\documentclass[10pt,conference,a4paper]{IEEEtran}
\usepackage{times,amsmath,epsfig}
\usepackage{epstopdf}

\usepackage{url}
\title{\textbf{HCMapper}: An interactive visualization tool to compare partition-based
flat clustering extracted from pairs of dendrograms}
\author{%
{Gautier Marti{\small $~^{\#*}$}, Philippe Donnat{\small $~^{\#}$}, Frank Nielsen{\small $~^{*}$}, Philippe Very{\small $~^{\#}$} } %  Removed for anonymous submission
{}
\vspace{1.6mm}\\
\fontsize{10}{10}\selectfont\itshape
$^{\#}$\,Hellebore Capital Management\\ %  Removed for anonymous submission
63 Avenue des Champs-\'Elys\'ees 75008 Paris FRANCE\\ %  Removed for anonymous submission
\,\\ 
$^{*}$\,Laboration d'Informatique de l'\'Ecole polytechnique\\ %  Removed for anonymous submission
LIX - \'Ecole polytechnique - 91128 Palaiseau Cedex FRANCE\\ %  Removed for anonymous submission
\\
\fontsize{9}{9}\selectfont\ttfamily\upshape
\fontsize{10}{10}\selectfont\rmfamily\itshape
\fontsize{9}{9}\selectfont\ttfamily\upshape
\,
}
\begin{document}
\maketitle

% INCLUDES COPYRIGHT NOTICE: one of three copyright notice should be included. Uncomment the appropriate line below, according to the authors affiliation:
\begin{figure}[b]
\parbox{\hsize}{\em
%information about the event:
%IEEE VCIP'15, Dec. 13 - Dec. 16, 2015, Singapore

%copyright notice: one of three copyright notices below should be included. Uncomment the appropriate line, according to the authors affiliation:
%000-0-0000-0000-0/00/\$31.00 \ \copyright 2015 IEEE.
%U.S. Government work not protected by U.S. copyright.
%???-?-????-????-?/10/\$??.?? \copyright 2015 Crown.
}\end{figure}

\begin{abstract}
We describe a new visualization tool, dubbed HCMapper, that visually helps
to compare a pair of dendrograms computed on the same dataset by displaying multiscale partition-based layered structures.
The dendrograms are obtained by hierarchical clustering techniques whose output reflects some hypothesis on the data and HCMapper
is specifically designed to grasp at first glance both whether the two compared hypotheses broadly agree and the data points on which they do not concur.
Leveraging juxtaposition and explicit encodings, HCMapper focus on two selected partitions while displaying coarser ones in context areas for
understanding multiscale structure and eventually switching the selected partitions.
HCMapper utility is shown through the example of testing whether the prices of credit default swap financial time series only undergo correlation.
This use case is detailed in the supplementary material at \url{www.datagrapple.com/Tech} as well as experiments with code on toy-datasets for reproducible research.
HCMapper is currently released as a visualization tool on the DataGrapple time series and clustering analysis platform at \url{www.datagrapple.com}.
\\[1\baselineskip]
\end{abstract}

% NOTE keywords are not used for conference papers so do not populate them
\begin{keywords}
Viewing algorithms; Model validation and analysis; Hierarchical clustering; Time series analysis
\end{keywords}

\section{Introduction}
Hierarchical clustering is a pervasive quantitative technique used in many fields leveraging exploratory analysis such as audio \cite{journals/taslp/EssidRD06} or image \cite{garcia2010levels} signal processing, social sciences, natural language processing \cite{ushioda1996hierarchical,glover2002inferring}, finance from real estate \cite{hepsen2012using} and stock \cite{tumminello2010correlation} portfolios analysis for investment, hedging or risk management purposes, to labour division amongst analysts \cite{yaros2013crowdsourced},
and foremost in phylogenetics where the hierarchy, the Tree of Life, is the gist \cite{blanchette2012inference,ailon2005fitting}.
Thus, it is paramount to be able to understand and compare hierarchical clustering structures. To do so, many approaches are available: as a first step, one can look for an isomorphism between trees which is decidable in linear time in the number of vertices \cite{IsoTree}; then, one can try to quantify how dissimilar they may be using, for instance, editing distance \cite{zhang1992editing} whose computation was shown to be an NP-complete problem, Robinson-Foulds distance \cite{robinson1981comparison} and its linear-time computation in the number of leaves \cite{day1985optimal}, or an index \cite{morlini2012dissimilarity} extending the comparison of partitions \cite{hubert1985comparing} to the whole hierarchies. However, the raw number yielded may be difficult to interpret, and does neither point out where differences appear nor what they consist in, failing to provide substantial information to the practitioner who may be interested by studying the cluster splits and the divergences between the two hierarchical structures.
Yet the object of study, namely hierarchical organization, is also popular to visualize data \cite{elmqvist2010hierarchical}: it represents information at different levels of detail, and is amenable to interactive browsing, e.g. users can drill up and down the hierarchy. Much work has been done for the last two decades to effectively visualize trees: Cone Tree \cite{robertson1991cone}, Treemaps \cite{johnson1991tree,balzer2005voronoi}, Hyperbolic Tree \cite{lamping1995focus+}, and many more \cite{jurgensmann2010poster}.
In comparison, less works have been led for representing and comparing multiple trees \cite{graham2010survey} despite its importance for field experts and practitioners who need to understand how dissimilar are clustering hierarchies produced by different clustering methods since these structural instabilities might point out objects of special interest.
The Hierarchical Clustering Mapper aims at finding them: it highlights clustering singularities between two models.
After reviewing related work, we explain in-depth design and features of the HCMapper before briefly showing its usefulness for testing two hypotheses on financial time series and finally discuss its current limitations.

\section{Related work}

In the visualization literature, tree comparisons have been adressed by leveraging the display space. Two trees can be displayed side by side \cite{TreeJuxtaposer}, one on top of the other \cite{CandidTree}, or the two facing each other \cite{VizHOD}. When dealing with (much) more than two trees to compare, \cite{TreeSet} and its extension \cite{hillis2005analysis} consider a treespace by using a distance between trees, namely Robinson-Foulds metric, and project, using multidimensional scaling, hundreds of trees corresponding to phylogenies on the Euclidean plane. The purpose is to look for consensus between models depicted by clusters of tree-points on the plane. By hovering over a cluster, they display the consensus tree associated to it, but then for finer comparison of the potentially several consensus trees produced they just display them side by side coming back to the problem of pair comparison using MacClade or Mesquite softwares, or TreeJuxtaposer \cite{TreeJuxtaposer} leveraging a focus+context approach to deal with large trees. Taking for granted that tree comparisons may require lots of display space, authors of \cite{isenberg2007interactive} developed an impressive tabletop framework to display trees under various representations and allow collaborative work of a team leaned on the tabletop to compare them. Others more interested in attributed trees, i.e. trees whose leaf nodes are associated with attributes, than in their topologies \cite{tu2007visualizing} opt for innovative treemap layouts, still displaying them side by side, but claiming that users can better understand the changes in the hierarchy and layout, and notice more quickly the color and size differences describing changes in the attributes. Alternatively, some decide to superimpose the two trees to primarily deal with uncertainty on some part of the trees \cite{CandidTree}, but authors of CandidTree mention that it can also be applied to show differences between hierarchical structures by giving the exemple of backup directory structures to grasp which files or folders were deleted or moved. But once again, the focus is on the tree topology and not on underlying partition-based clustering. Designs mentioned fall into the general taxonomy of visual designs for comparison 
\cite{gleicher2011visual}, they involve juxtaposition or superposition. The remaining category, namely explicit encoding, has been used in tanglegrams \cite{nollenburg2009drawing, galili2015dendextend} which are pair of trees whose leaf sets are in one-to-one correspondence and whose matching leaves are connected by inter-tree edges. Actual comparison is based on the measurement of minimum edge entanglements connecting the leaves from the two hierarchies, yet minimizing the entanglements, a two-sided crossing minimization problem (2SCM), is an NP-hard problem, and thus results are heuristics-dependent.
An extension of the tanglegram is presented in \cite{VizHOD} focusing on comparing hierarchically organized data in the context of software systems by using an edge-bundling technique which highlight splits, joints and relocations of subhierarchies. The latter bears much resemblance to our work at first sight but our diverging goals have led to different designs, ours focusing more on consensus and outliers between two partitions at a time rather than providing a global overview of the differences and similarities between the two hierarchies in a single picture.

\section{Hierarchical Clustering Mapper explained}

In this section, we detail the design of HCMapper by first describing the mapping from input data to partition layers, and then how these layers are displayed to highlight moot points from the dataset.

%In this section, we pose the problem we want to solve thanks to a visualisation, and then introduce the different aspects of the HCMapper visualization that is a first step in the literature to solve this specific task of importance for practitioners, at least in the financial industry.% tool through its features, its formal definition and computational complexity for building and displaying it. Then, we present the elementary open source blocks available to reproduce it effectively.

%\subsection{Adressed problem: Who are the deviants?}

%look for divergences/outliers in dendrograms which defines partitions of the data; each dendrogram is an hypothesis of the dataset; the two dendrograms are defined on the same set of labeled points thus we can find correspondence between layers which are partitions;

%bigger the edge, stronger the consensus
%otherwise, How strong the consensus?

\subsection{Construction and visualization layout}

\begin{figure}[htb]
  \centering
  % the following command controls the width of the embedded PS file
  % (relative to the width of the current column)
  \includegraphics[width=.95\linewidth]{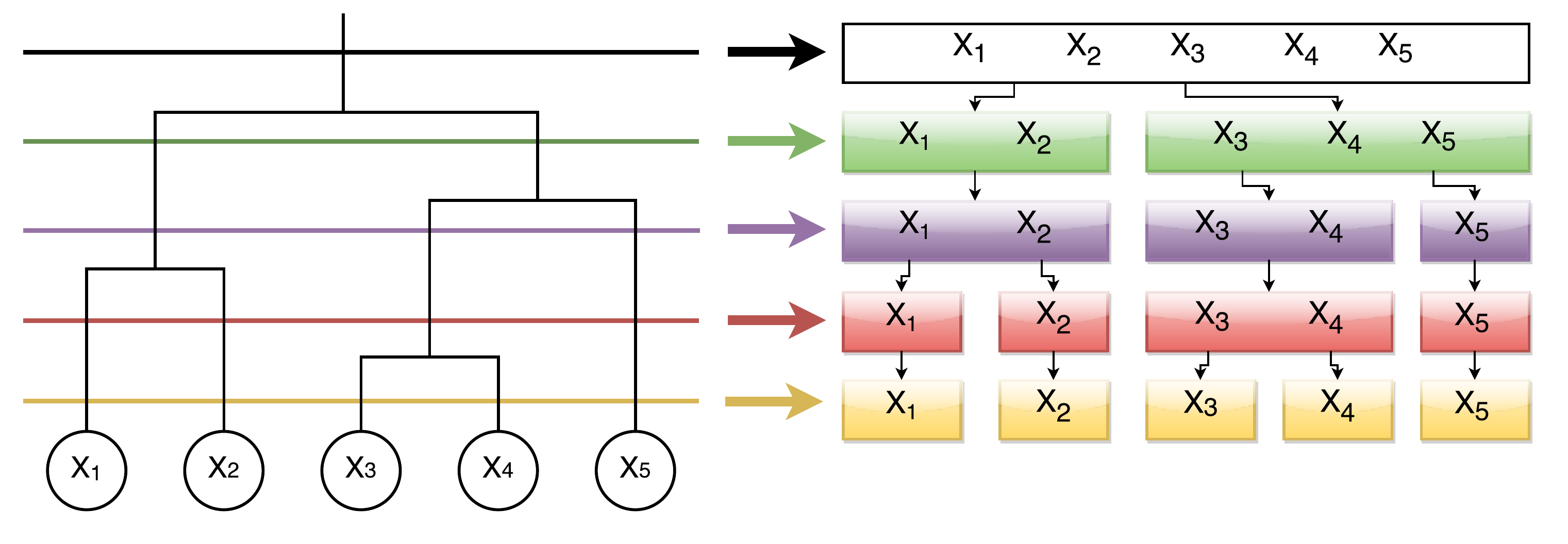}
  \caption{Extracting flat partition-based clustering from a dendrogram and transforming it into a tree of clusters; all clusters at a given depth in this tree form a partition over the dataset $\mathcal{X} = \{x_1,x_2,x_3,x_4,x_5\}$.}
  \label{dend2layers}
\end{figure}

We start from a dataset $\mathcal{X} = \{x_1,\ldots,x_n\}$. We obtain a dendrogram on $\mathcal{X}$ using a hierarchical clustering algorithm whose output reflects an hypothesis over $\mathcal{X}$. For comparing two hypotheses over $\mathcal{X}$, we begin by building two such dendrograms. Then, we extract from each dendrogram all possible flat partitions over $\mathcal{X}$, thus transforming each one into a tree whose vertices at a given depth define a partition over $\mathcal{X}$, partitions ranging from the coarsest one at the root to the finest one at the leaves as illustrated in Figure~\ref{dend2layers}.

Bonds between the two tree vertices are explicitly encoded as links representing whether their associated cluster intersects or not. To avoid decontextualization using only explicit encoding design between two partitions, we also use juxtaposition by displaying coarser layers from each tree to understand the emergence of the two studied partitions, and a multiple view interactive system to switch the studied partitions. To avoid cluttering in this hybrid approach, we can activate a fisheye effect to read textual information if needed. Since cluster size is a significant information of clustering, we display it through its associated vertex whose size is proportional to the ratio of the cluster size over the dataset size. Even more important to our task, understanding clusters intersection: intersection size is encoded through the size of the edge that bonds them, wider is the edge relatively to the cluster size, stronger the consensus; intersection content can be displayed by hovering over the edge. To display this graph, we can leverage the D3.js \cite{bostock2011d3} Sankey, a highly customizable chart, which is amenable for adding application-oriented features such as the ones recommended in \cite{guerra2013visualizing} for visualizing dynamic hierarchies. Yet inspiring we think this work poorly handles outliers using a reporting tool which merely lists them in textual form. We believe that HCMapper is more suited to this specific task.
Concerning time complexity for building the visualization from scratch on $\mathcal{X} = \{x_1,\ldots,x_n\}$, it requires $O(n^2 \log n)$ for applying agglomerative hierarchical clustering algorithm, then for transforming the dendrograms obtained it costs $O(n^2)$, and finally for displaying $V$ vertices $O(V^3)$. 

\subsection{HCMapper reading}

\begin{figure}[htb]
  \centering
  % the following command controls the width of the embedded PS file
  % (relative to the width of the current column)
  \includegraphics[width=.99\linewidth]{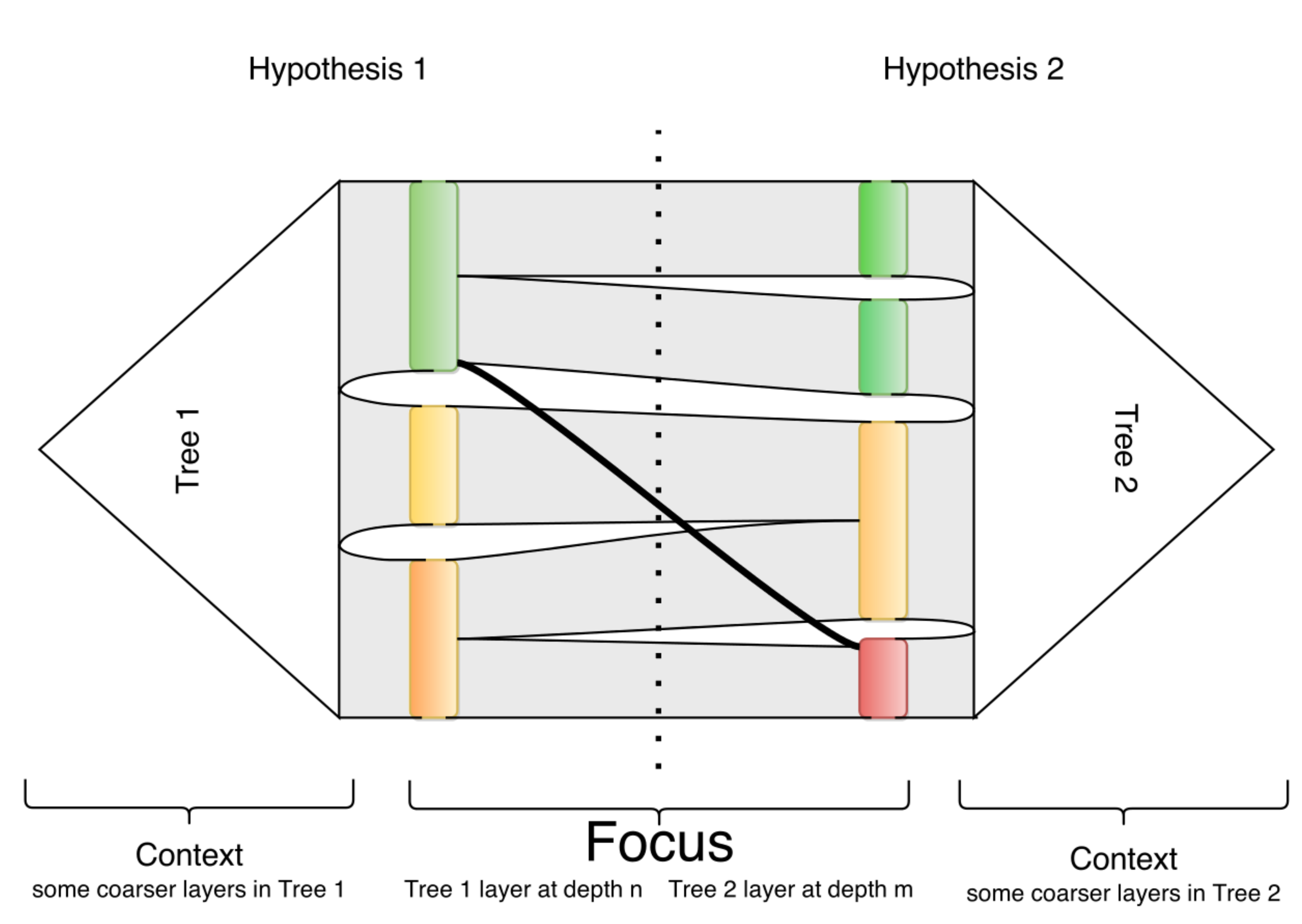}
  \caption{Two hypotheses are compared through the dendrograms which were transformed into two trees of partition-layers; note the diverging edge from the green cluster to the red one which highlights a moot point of special interest for experts.}
  \label{compare_dend}
\end{figure}

\begin{figure}[htb]
	\includegraphics[width=.99\linewidth]{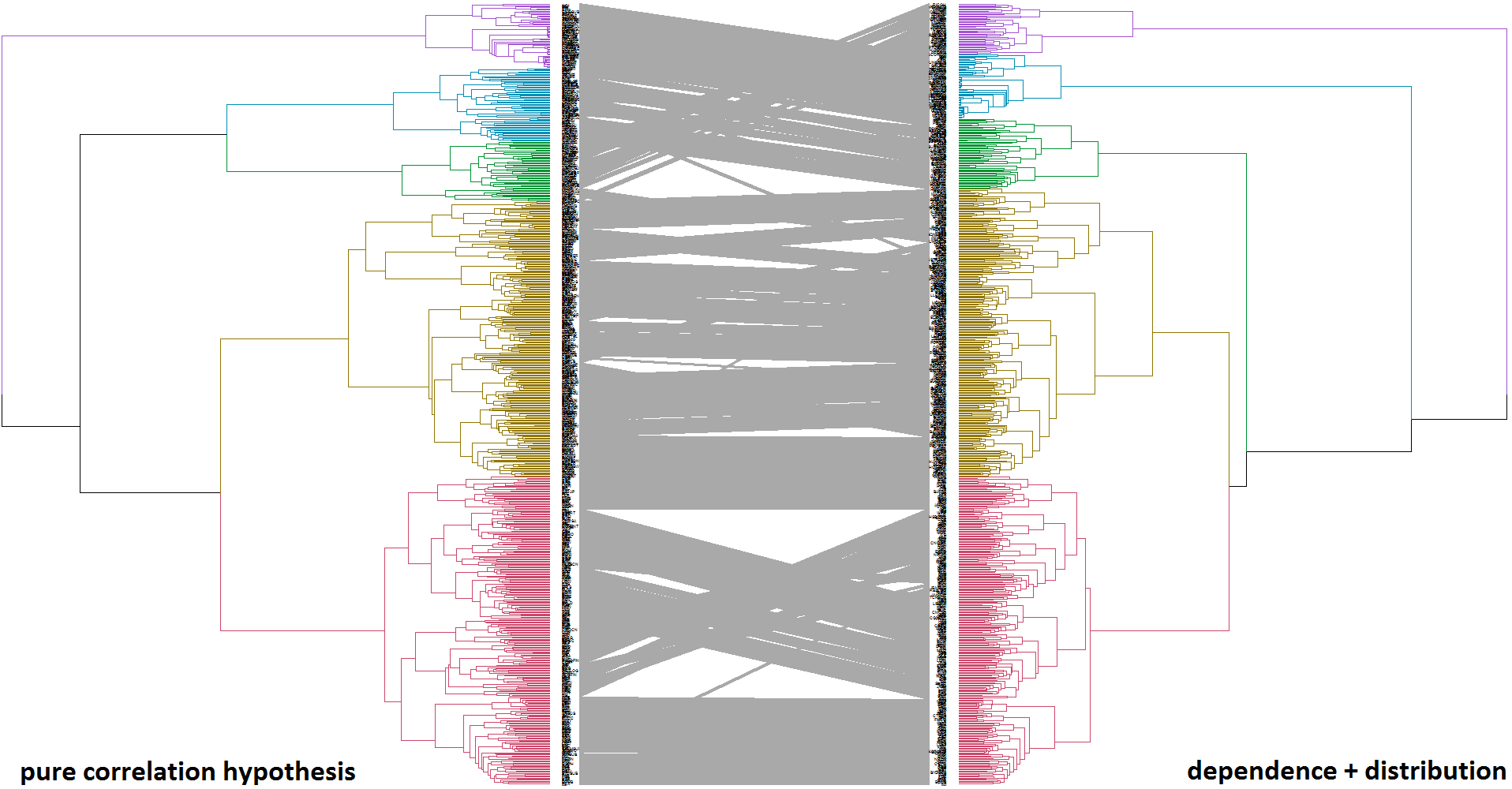}
      \caption{\label{fig:Tanglegram1Star}
            Left: two partitions extracted from a dendrogram for pure correlation hypothesis; Right: two partitions extracted from a dendrogram for dependence+distribution hypothesis; Consensus or divergence is much harder to grasp in the dendextend tanglegram \cite{galili2015dendextend} than in HCMapper.}
\end{figure}

The HCMapper graph is depicted in Figure~\ref{compare_dend}. There are three main areas: two contexts, and a focus. The left and right contexts (Tree 1 and Tree 2 in the picture) represent some of the successive coarser layers of the left and right inner layers (colored layers in the image) to be compared in the focus area. The main goal is to understand how these two partitions are related. In Figure~\ref{compare_dend}, we compare a partition at depth $n$ obtained from a dendrogram reflecting hypothesis 1, and a partition at depth $m$ obtained from the alternative dendrogram reflecting hypothesis 2. Looking at the left partition in the focus area, we can see that it is composed of three clusters (green, yellow, orange) of approximately equal size. The right partition consists in four clusters (light green, deep green, light orange and red). We can observe that the left green cluster is mapped onto the light and deep green clusters from the right partition which consist in a refinement.
However, right partition is not a strict refinement of the left partition as it can be seen by looking at the right light orange cluster. Indeed, this one is composed of the left yellow cluster and a part of the left orange cluster, the latter evenly splitting into the right light orange cluster and the red one.
But, what is the most important to notice here, and which is visually blatant, is the small edge diverging from the bulk of edges linking the left green cluster to the two right green clusters. This singular edge links an element from the left green cluster to the corresponding same labeled element yet belonging to the red cluster in the right partition. This actually can be considered as an outlier or a point of special interest since the two hypotheses broadly agree but on this point.
Notice that finding so readily this special point would be much harder by looking through clustering textual results or by inspecting common visual comparison between dendrograms such as the tanglegram displayed in Figure~\ref{fig:Tanglegram1Star} which is implemented in the dendextend R package.

\section{Application to financial time series analysis}

\begin{figure}[htb]
	\includegraphics[width=.99\linewidth]{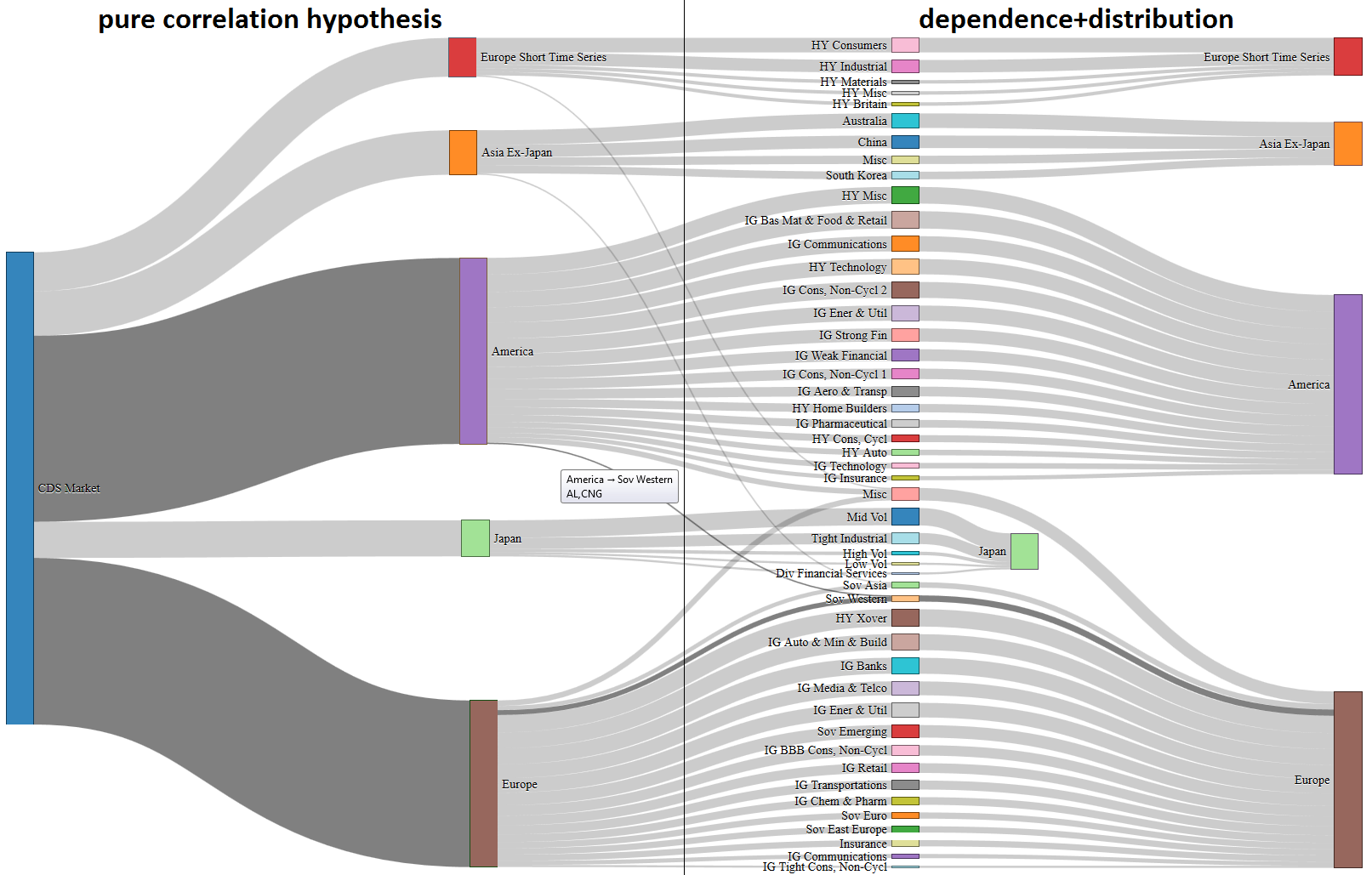}
      \caption{\label{fig:Theta1Star}
            Left: two partitions extracted from a dendrogram for pure correlation hypothesis; Right: two partitions extracted from a dendrogram for dependence+distribution hypothesis. We can readily conclude that correlation is paramount for clustering prices. By hovering over the thin edge diverging from the ``America" cluster in the pure correlation hypothesis, we can read AL (Rio Tinto) and CNG (AT\&T) on the tooltip. These American companies are clustered with high quality European government debt assets (e.g. Norway, Sweden, Denmark), all of them having a daily variation distribution characterizing illiquid products.}
\end{figure}

Quantitative analysts and decision-makers have noticed that on some financial markets such as stocks, bonds and vanilla derivatives, time series of assets prices tend to cluster according to their pairwise correlation \cite{tumminello2010correlation}. But, is correlation really the only explanatory factor? By assuming a random walk model for prices, correlation and distribution amount for the whole information \cite{sklar1959fonctions} in these time series. To answer this question, we state two hypotheses: there is only correlation in the data (H1); there is correlation and distribution in the data (H2). Then, we compute a dendrogram reflecting H1 and another for H2 before building the HCMapper graph displayed in Figure~\ref{fig:Theta1Star}. At once, one can notice that clusters from H1 hypothesis are broadly in a one-to-one correspondence with clusters from the H2 hypothesis, but a few outliers highlighted by thin diverging edges. Thus, thanks to the HCMapper visualization, we can conclude that correlation is the main explanatory factor for the clustering of prices time series since only few clusters are modified by adding the distribution information. This can be explained by market microstructure: market makers are specialized and cover specific sectors, thus adding correlation between prices  which are already influenced by common macroeconomic factors.
Yet, the few moot points found are of paramount significance for experts as they may correspond to assets whose price variations undergo heavy-tailed distribution or suffer from illiquidity, therefore particular attention should be given to these assets while performing a risk analysis. For a thorough treatment of this section, refer to \cite{donnat2016toward,marti2015proposal} and \url{www.datagrapple.com/Tech}.

%-------------------------------------------------------------------------
\section{Conclusions and future work}

HCMapper is an efficient tool for detecting moot points in a dataset with respect to several hypotheses described by different hierarchical clusterings. Its time complexity in $O(n^2 \log n + V^3)$, where $n$ is the number of points and $V$ the number of vertices to display,  makes it unusable for very large datasets (cost of hierarchical clustering) or large display graphs (cost of positioning vertices).
To our opinion, HCMapper is a promising tool for comparing dendrograms, provided it undergoes adaptations to fit to one's needs. Concerning its evolution, we plan to extend HCMapper scope and ergonomy by using an information-geometric approach \cite{Nock:2013:ILM:2542355.2542378} allowing for multiple foci+contexts to simultaneously zoom-in on several interesting cluster splits. 
HCMapper is publicly available and in daily use: this visual tool was  
integrated in the DataGrapple project platform (\url{www.datagrapple.com}) where anyone can test its usefulness on credit default swaps historical time series data.
We have released both implementation and use case, and hope it will be helpful for biologist researchers as well.

%\vfill\eject

\bibliographystyle{IEEEtran}

\bibliography{IEEEabrv,IEEEexample}

\end{document}